# A Neuro-Fuzzy Model for Function Point Calibration

Wei Xia[1], Danny Ho[2], Luiz Fernando Capretz[3]

[1]HSBC Bank Canada, IT Department, Vancouver, BC, Canada
[2]NFA Estimation Inc., London, ON, Canada
[3]University of Western Ontario, Dept. Electrical & Computer Engineering, London, ON, Canada

**Abstract:** The need to update the calibration of Function Point (FP) complexity weights is discussed, whose aims are to fit specific software application, to reflect software industry trend, and to improve cost estimation. Neuro-Fuzzy is a technique that incorporates the learning ability from neural network and the ability to capture human knowledge from fuzzy logic. The empirical validation using ISBSG data repository Release 8 shows a 22% improvement in software effort estimation after calibration using Neuro-Fuzzy technique.

## 1. Introduction

Software projects are infamous for exceeding their original estimates. Effort overruns usually lead to cost overruns, which in turn may lead to fines for the contractors for late delivery; schedule overruns may result in loss of business [1]. Researchers have directed their work at improving estimation accuracy using systematic models that based on a variety of measures of software size, such as lines of code (LOC) [2] and Function Point (FP) [3].

FP is an ideal software size metric to estimate cost since it can be obtained in the early development phase, such as requirement, measures the software functional size from user's view, and is programming language independent [4]. To achieve more accurate estimation, we suggest the concepts of calibration detailed in the next section.

Neural network technique is based on the principle of learning from historical data. The neural network is trained with a series of inputs and desired outputs from the training data set [5]. Once the training is complete, new inputs are presented to the neural network to predict the corresponding outputs. Fuzzy logic is a technique to make rational decisions in an environment of uncertainty and imprecision. It is rich in representing human linguistic ability with the terms such as fuzzy set, fuzzy rules [6], [7]. Once the concept of fuzzy logic is incorporated into the neural network, the result is a Neuro-Fuzzy system that combines the advantages of both techniques [8], [9]. This technique is found appropriate to calibrate FP as proved by the validation results.

Yau and Tsoi [10] introduce a fuzzified FP analysis model to help software size estimators to express their judgment and use fuzzy B-spline membership function to derive their assessment values. The weak point of this work is that they used limited in-house software to validate their model, which brings a great limitation regarding the validation of their model. Lima, Farias and Belchior [11] also proposed the use of concepts and properties from fuzzy set theory to extend FP analysis into a fuzzy FP analysis, a prototype that automates the calculation of FPs using the fuzzified model was created, but again the calibration was done using a small database comprised of legacy systems developed mainly in Natural 2, Microsoft Access and Microsoft Visual Basic, which compromises this work's generality.

Al-Hajri et al. [12] establish a new FP weight system using artificial neural network. Like our work, in order to validate their model, they also used the data set provided by the International Software Benchmarking Standards Group (ISBSG). In their research, tables gathered with the training methods from neural networks replaced the original complexity table. Their results are quite accurate, although the correlation is still unsatisfactory with MMRE over 100%, which originates from the wide variation of data points with many outliers.

**Concept of Calibration**

A neuro-Fuzzy FP calibration model that incorporates the learning ability from neural network and the ability to capture human knowledge from fuzzy logic is proposed and further validated.

Calibrate to fit specific application

In FP counting, each component, such as Internal Logical File (ILF), is classified to a complexity determined by its associated files numbers, such as Data Element Types (DET), Record Element Types (RET) [4] as listed in Table 1. Such complexity classification is easy to operate, but it may not fully reflect the software complexity under the specific software application. For example, Table 2 shows a software project with three ILFs: A, B and C. According to the complexity matrix, A and B are classified as having the same complexity and are assigned the same weight value of 10. However, A has 30 more DET than B and is certainly more complex, but they are now assigned the same complexity. Also, B is classified as average and assigned a weight of 10 while C is classified as low and assigned a weight of 7. B has only one more DET than C and the same number of RET as C, but B has been assigned three more weight units than C. There is no smooth transition boundary between two classifications. Processing the number of FPs component associated files such as DET, RET using fuzzy logic can produce an exact complexity degree.

Table 1: Observations of FP Complexity Classification.

|  | ILF A | ILF B | ILF C |
|---|---|---|---|
| **DET** | 50 | 20 | 19 |
| **RET** | 3 | 3 | 3 |
| **Complexity** | Average | Average | Low |
| **Weight Value** | 10 | 10 | 7 |

Table 2: ILF Complexity Matrix.

| ILF | DET | | |
|---|---|---|---|
| RET | 1-19 | 20-50 | 51+ |
| 1 | Low | Low | Avg |
| 2-5 | Low | Avg | High |
| 6+ | Avg | High | High |

Calibrate to reflect industry trend

The weight values of Unadjusted Function Points - UFP (Table 3) are said to reflect the functional size of software [3]. They were determined by Albrecht in 1979 based on the study of 22 IBM Data Processing projects. Since 1979, software development has been growing steadily and is not limited to one organization or one type of software. Thus, there is a need to calibrate these weight values to reflect the current software industry trend. The International Software Benchmarking Standard Group (ISBSG) maintains a large empirical project data repository. ISBSG data repository Release 8 contains 2,027 projects, which cover a broad range of project types with 75% of the projects being less than 5 years old. Learning UFP weight values from ISBSG Release 8 using neural network calibrates FP to reflect the current software industry trend.

Table 3: UFP Weight Values.

| Component | Low | Average | High |
|---|---|---|---|
| External Inputs | 3 | 4 | 6 |
| External Outputs | 4 | 5 | 7 |
| External Inquiries | 3 | 4 | 6 |
| Internal Logical Files | 7 | 10 | 15 |
| External Interface Files | 5 | 7 | 10 |

Calibrate to improve cost estimation

The significant relationship between the software size and cost has been recognized for a long time. In the classical view of cost estimation process (Figure 1), the outputs of *effort* and *duration* are estimated from software *size* as the primary input and a number of *cost factors* as the secondary inputs. There are mainly two types of software size metrics: Source Lines of Code (SLOC) and FP. SLOC is a natural artifact that measures software physical size, but it is usually not available until the coding phase and difficult to have the same definition across different programming languages. FP is an ideal software size metric to estimate cost since it can be obtained in the early development phase, such as requirement, measures the software functional size, and is programming language independent [4]. Calibrating FP incorporates the historical information and gives a more accurate view of software size. Hence more accurate cost estimation comes with a better software size metric.

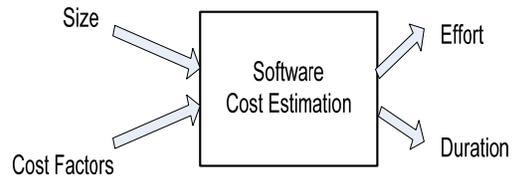

**Figure 1: Classical View of Cost Estimation Process.**

## 2. Neuro-Fuzzy Calibration Approach

We propose an approach to calibrate FP using Neuro-Fuzzy technique. The model overview and two parts of the model: fuzzy logic part and neural network part are described here. The empirical validation is provided in the next section. The block diagram shown in Figure 2 gives an overview of our approach. The project data provided by ISBSG [13] is imported to extract an estimation equation and to train the neural network. The estimation equation is extracted from the data set by statistical regression analysis. Fuzzy logic is used to calibrate FP complexity weights to fit specific application and to reflect the current software trend. The validation results show that the calibrated FP weights have better estimation ability t original.

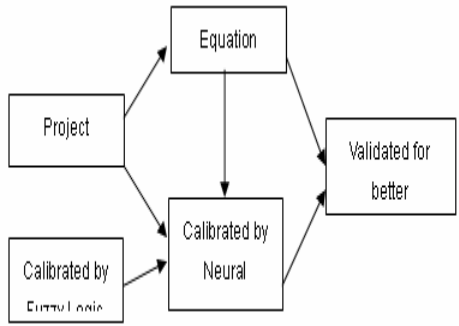

**Figure 2: Block Diagram of Neuro-Fuzzy Approach.**

Fuzzy Logic Part

The fuzzy logic part calibrates the FP complexity degree to fit the specific application. A fuzzy logic system (shown in Figure 3) is constructed based on the fuzzy set, fuzzy rules and fuzzy inference. The input fuzzy sets are to fuzzify the component associated file numbers and the output fuzzy set are to fuzzify the complexity classification. The fuzzy rules are defined in accordance with the original complexity weight matrices. The fuzzy inference process using the Mamdani approach [14] is applied based on the fuzzy sets and fuzzy rules.

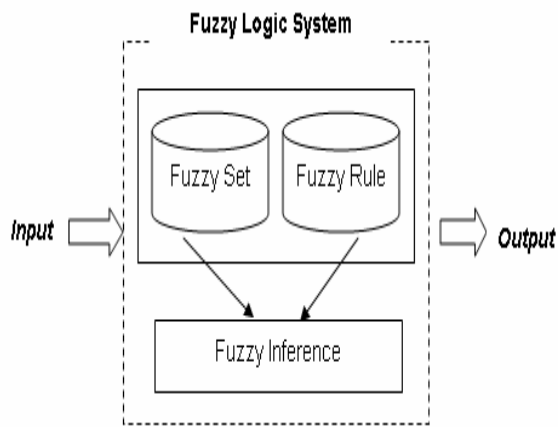

**Figure 3: Fuzzy Logic Syste**

Neural Network Part

The neural network part is aiming at calibrating FP to reflect the current software industry trend. By learning from ISBSG data repository, this part is believed to achieve the calibration goal.

First, we need to come up with an equation that can estimate the software cost in work effort; and is then used in the neural network training as an activation function. Similar estimation equations are proposed in [15] and [16], but we chose to construct a new equation based on the ISBSG Release 8 data repository in order to meet the following requirements:

**Reliability**: The equation should be built on a reliable data set to avoid the situation described in the old saying "garbage in garbage out". Thus, certain filters must be applied to the raw data set for quality purposes. The equation must be derived by sound statistical regression analysis.

**Extendability**: The equation is derived based on the ISBSG data repository. However, the equation should be able to be applied to other FP oriented data repositories as well. Hence, the elements in the equation should not be unique in the ISBSG data repository. Also, since ISBSG Release 8 does not provide data points on well-defined cost drivers like the Constructive Cost Model (COCOMO) factors [17], so the equation should have room to be extended to include cost drivers for future investigations.

**Integrality**: To arrive at an equation that estimates the software cost in work effort from FP, many predictors can be taken into account, but to ensure integrity only the essential predictors should be included in the final equation.

In order to reach a reasonable conclusion, the raw ISBSG data set is filtered by several criteria recommended by ISBSG [18]. A subset of 184 projects is obtained of which the quality rating is A or B, the counting method is IFPUG which excludes other counting methods such as COSMIC FFP [19] and Mark II [20], the effort resource is recorded at level one (development team), the development type is new development or re-development, the 15 Unadjusted Function Point (UFP) breakdowns and 14 General System Characteristics rating values are available.

The data obtained after the applying filtering criteria were transformed to satisfy the assumptions for regression analysis. The positive relationship between the work effort and size has been reported in [15], [16], and [21]. FP is a functional software size that is reached by multiplying UFP and the VAF. However, the definition of VAF that is supposed to reflect the technical complexity has been criticized for overlapping and being out of date [22], [23]. The unadjusted FP has been standardized as software unadjusted functional size measurement through the International Organization for Standardization (ISO) [24], but VAF has not. The ISBSG Release 8 data field description document recommends using the field of "Normalized Work Effort" as a project work effort. Thus, we have chosen the unadjusted FP as the software size and normalized work effort as the work effort in statistical analysis.

The statistical regression analysis assumes that the underlying data are normally distributed. However, the histograms of effort and size (Figures 4(a) and 4(b)) show that they are not distributed normally but are highly skewed. To approximate a normal distribution, we apply a logarithmic transformation on these variables to make the large values smaller and to bring the data closer together. It is observed in the histograms of *ln* UFP and *ln* Work Effort (Figures 4(c) and 4(d)) that the transformed variables are approximately normally distributed. The relationship between the work effort and size is visualized, using two-dimensional graphs as shown in Figures 5(a) and 5(b), before and after logarithmic transformation respectively. An obvious positive linear relationship between effort and size after logarithmic transformation is observed.

The neural network is constructed to receive 15 UFP breakdowns as inputs to give the work effort as the desired output. A back-propagation learning algorithm [25] is conducted in order to minimize the prediction difference between the estimated and actual efforts. An effort estimation equation is extracted based on the data subset using statistical regression analysis. The equation in the form of:

$$\textit{Effort} = A \cdot \textit{UFP}^{\,B} \qquad \text{Equation (1)}$$

is achieved with the help of the statistical software SPSS v12 [26].

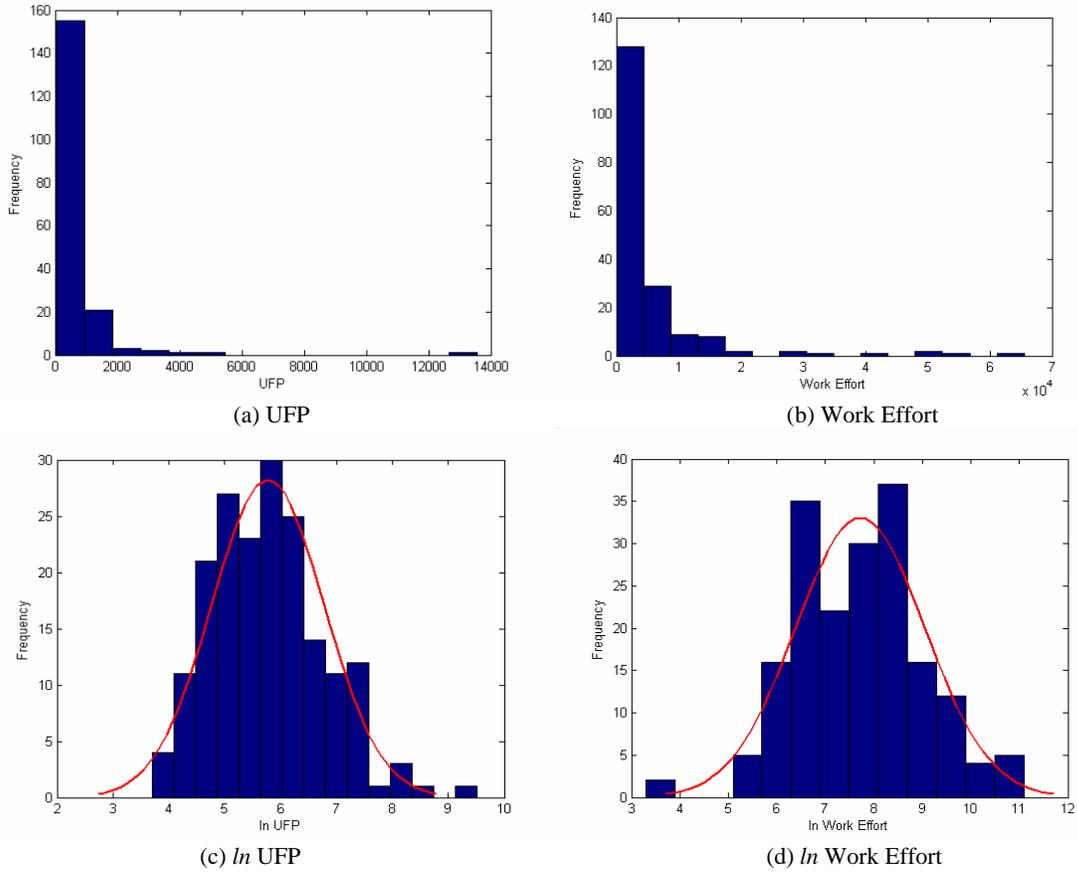

(a) UFP  (b) Work Effort
(c) *ln* UFP  (d) *ln* Work Effort

**Figure 4: Histograms of UFP and Work Effort**

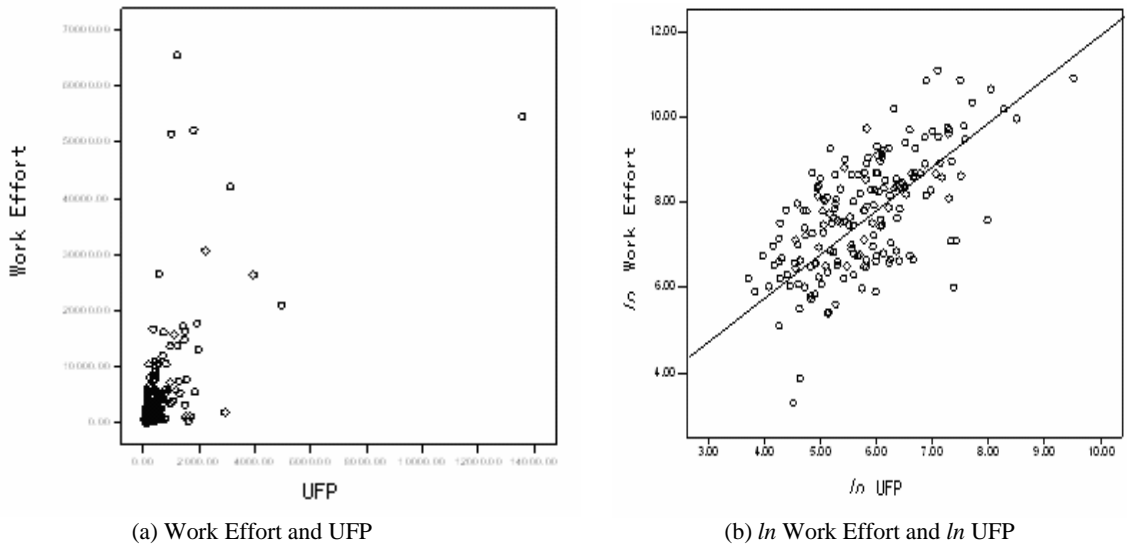

(a) Work Effort and UFP  (b) *ln* Work Effort and *ln* UFP

**Figure 5: Graphs of UFP and Work Effort**

Certain post regression analysis was done to check the regression validity. Regression models are based on the assumptions that the residuals are normally distributed, independent, with a constant variance and a mean value that equals zero. Residuals refer to the differences between the estimated and the desired values. The residuals are plotted in Figure 6 as a histogram and they appear to be approximately normally distributed, with a mean value of zero. Therefore, the assumptions for statistical regression analysis are satisfied and we can conclude that our equation is valid.

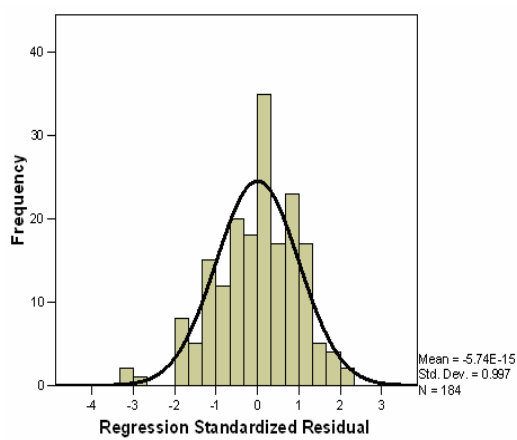

**Figure 6: Histogram of Residuals**

Though of simple form, Equation 6 satisfies the conditions listed in the beginning of the section. It is derived from the reliable filtered data set and analyzed by reliable statistical procedure that includes logarithmic transformation, statistical regression, and post regression analysis. It does not include any special ISBSG repository parameters; thus it can estimate FP-oriented projects and can be extended to include cost drivers for future works. It contains UFP as the only predictor and excludes VAF, a parameter receiving much criticism, thus ensuring the integrity.

## 3. Validation Results

Five experiments were conducted to validate our Neuro-Fuzzy approach. For each experiment, the original data set (184 projects) was randomly separated into 100 training data points and 84 test data points. The outliers are the abnormal project data points with large noise that may distort the training result. Thus, we used the training data set excluding the outliers for calibration, but used the rest of the data points for validation [27]. The average calibrated UFP weight values obtained from five experiments are listed in Table 4, and the original weight values are given as comparison.

The validation results of the five experiments are assessed by Mean Magnitude Relative Error (MMRE) for estimation accuracy. MMRE is defined as:

For $n$ projects:

$$MMRE = \frac{1}{n}\sum_{i=1}^{n}(|Est - Actual|/Actual)$$

Table 5 lists the validation results of the five experiments where "Improvement %" is the MMRE improvement in percentage for each experiment. Based on the MMRE assessment results, an average of 22% cost estimation improvement has been achieved with the Neuro-Fuzzy calibration approach. The MMRE after calibration is over 100% which is still relatively large and is due to the absence of well-defined cost drivers like COCOMO [17] factors.

The validation results of the five experiments are also assessed by Prediction at level $p$ (PRED) criteria, i.e., $PRED(p) = k/N$, where $N$ is the total number of projects, $k$ is the number of projects with absolute relative error of $p$. Four PRED criteria are assessed here, namely Pred 25, Pred 50, Pred 75 and Pred 100. Table 6 lists the Pred assessment result where the overall performance is improved.

## 4. Conclusion

The Neuro-Fuzzy approach to calibrate FP is validated with the empirical data repository (ISBSG Release 8). The experimental validation results show a 22% improvement in software cost estimation and demonstrate that FP need calibration and can be calibrated.

The fuzzy logic part of the model calibrates the FP complexity weights to fit the specific application context. The neural network part of the model calibrates the UFP weight values to reflect the current software industry trend. The combined neuro-fuzzy technique calibrates FPfor better cost estimation.

| Component | Low | | Average | | High | |
|---|---|---|---|---|---|---|
| | Original | Calibrated | Original | Calibrated | Original | Calibrated |
| External Inquiries | 3 | 0.9 | 4 | 2.2 | 6 | 4.7 |
| External Outputs | 4 | 3.3 | 5 | 4.6 | 7 | 6.2 |
| External Inquiries | 3 | 1.8 | 4 | 2.9 | 6 | 5.4 |
| Internal Logical Files | 7 | 5.4 | 10 | 9.8 | 15 | 14.9 |
| External Interface Files | 5 | 4.6 | 7 | 6.9 | 10 | 10 |

Table 4: Calibrated UFP Weight Values.

Table 5: MMRE Validation Result.

| | Exp.1 | Exp.2 | Exp.3 | Exp.4 | Exp.5 |
|---|---|---|---|---|---|
| **MMRE Original** | 1.38 | 1.58 | 1.57 | 1.39 | 1.42 |
| **MMRE Calibrated** | 1.10 | 1.28 | 1.17 | 1.03 | 1.11 |
| Improvement % | 20% | 19% | 25% | 26% | 22% |
| **Average Improvement %** | **22%** | | | | |

Table 6: PRED Validation Results.

| | Average Original | Average Calibrated | Average Improvement |
|---|---|---|---|
| **Pred 25** | 13% | 12% | -1% |
| **Pred 50** | 23% | 27% | 4% |
| **Pred 75** | 40% | 46% | 6% |
| **Pred 100** | 60% | 67% | 7% |